# Dynamic SLA Negotiation using Bandwidth Broker for Femtocell Networks

Mostafa Zaman Chowdhury*, Sunwoong Choi*, Yeong Min Jang*, Kap-Suk Park**, and Geun Il Yoo**
*Kookmin University, Korea
**Korea Telecom (KT), Korea
*{mzceee, schoi, yjang}@kookmin.ac.k

*Abstract*— Satisfaction level of femtocell users' depends on the availability of requested bandwidth. But the xDSL line that can be used for the backhauling of femtocell traffic cannot always provide sufficient bandwidth due to the inequality between the xDSL capacity and demanded bandwidth of home applications like, IPTV, PC, Wi-Fi, and others. A Service Level Agreement (SLA) between xDSL and femtocell operator (mobile operator) to reserve some bandwidth for the upcoming femtocell calls can increase the satisfaction level for femtocell users. In this paper we propose a SLA negotiation procedure for femtocell networks. The Bandwidth Broker controls the allocated bandwidth for femtocell users. Then we propose the dynamically reserve bandwidth scheme to increase the femtocell user's satisfaction level. Finally, we present our simulation results to validate the proposed scheme.

*Keyword*S— xDSL, FAP, femtocell, QoS, Bandwidth Broker, SLA, and bandwidth.

## I. Introduction

Home based wireless networks will be the key element to support large demand of multimedia traffics for IMT-Advanced networks. Thus 3GPP specified *Home NodeB (HNB)* [1] or *Femto Access Point (FAP)* that uses existing xDSL or other cable line to connect the FAP to core network is a strong candidate for home networks to support voice and data services in the home environment. The basic connection of femtocell access point FAP to core network is shown in Figure 1. xDSL is a generic abbreviation for the many flavors of DSL or Digital Subscriber Line technology. This is the technology that is used between a customer's premises and the telephone companies, enabling more bandwidth over the already installed copper cabling than users have traditionally had [2]. Typically, the download speed of consumer DSL services ranges from 256 kbps to 50 Mbps, or more depending primarily on the equipment used, distances involved, cabling quality, encoding techniques, frequency spectrum available and even to some degree, end system configurations.

The availability of requested bandwidth by femtocell users is the key element to ensure the *Quality of Service (QoS)*. In the home environment same xDSL line is used for several non-femtocell services like Wi-Fi, IPTV, personal computer (PC), and etc. simultaneously with the femtocell users. So, for heavy internet traffic condition, it can not be ensured to support the femtocell users with the sufficient bandwidth. Most of the non-femtocell services are not delay sensitive and can be assumed as background traffic whereas most of the femtocell services like voice and video calls are delay sensitive. Thus, the quality of voice and other real-time calls degrade significantly as well as users' satisfaction level also decreases. Limitation of xDSL backhaul capacity can degrade the QoS in terms of packet loss, jitter, delay, and throughput.

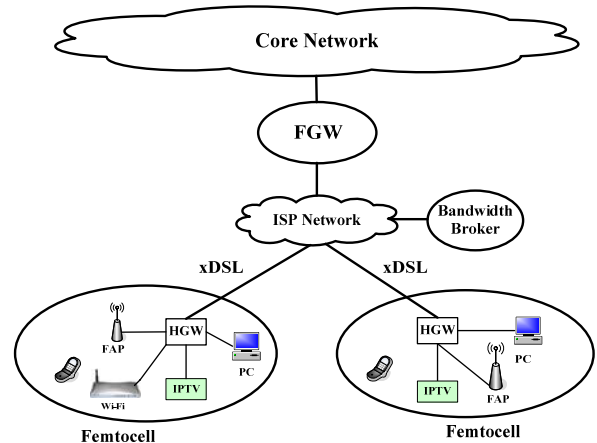

**Figure 1.** Connection of multiple devices using same ISP network

A *Service Level Agreement (SLA)* [3], [9] negotiation between the xDSL operator and FAP operator (mobile operator) can provide sufficient bandwidth for femtocell users. The average bandwidth requirement for 6 users' femtocell is not more than 500 kbps. But for the inequality between xDSL capacity and bandwidth demanded by non-femtocell services at home, some requested femtocell calls are blocked and some ongoing femtocell calls are dropped or quality degraded. Due to the SLA negotiation, xDSL can reserve sufficient bandwidth for femtocell calls. The Bandwidth Broker [4], [5] can be used to allocate bandwidth for femtocell users. The BB makes a policy access, resource reservation and admission control decisions for the SLA framework. From the previous history of requested bandwidth by femtocell calls, Bandwidth Broker (BB) reserves bandwidth for femtocell calls. The BB calculates the average requested bandwidth from the previous call history for a certain amount of time. Then the BB allocates this amount of bandwidth for femtocell users from the xDSL capacity. The remaining bandwidth can be used by other

This research was supported by the MKE (Ministry of Knowledge and Economy), Korea, under the ITRC (Information Technology Research Center) support program supervised by the IITA (Institute of Information Technology Assessment) (IITA-2009-C1090-0902-0019).

non-femtocell applications in the home. As BB considers only a certain period of recent call history, there is not much more difference between the dynamically changed reserve bandwidth and the requested bandwidth by femtocell users. Thus the bandwidth utilization of xDSL does not decrease significantly but satisfaction level of femtocell user's increases significantly.

This paper is organized as follows. Section II provides related background studies about SLA, BB, and Home Gateway. The proposed dynamic bandwidth reservation scheme and SLA framework are presented in Section III. In Section IV, the simulation results for the proposed scheme are presented to validate the scheme. Finally, we give our conclusion in Section V.

## II. Background

Mobile operators use xDSL or other broadband cable networks for the traffic backhauling of femtocell users. ISP operator and mobile operator may not same. The end user may use different operator. Hence different operators are involved with the integrated femtocell/macrocell networks. Due to these heterogeneous operators, there should have inter-operator agreement to provide end-to-end QoS for femtocell users. In a typical home environment, multiple equipments, such as FAP, laptop, PC, IPTV or a Wi-Fi, could share the same xDSL connection using *Home Gateway (HGW)*. A good voice quality requires low latency and less packets drop, but none of them are guaranteed by current ISP networks. Quality of the backhaul connection typically depends on xDSL capacity, overall backhaul network load, and bandwidth management. Voice traffic from the femtocell competes with other multiple equipments to send data through broadband connection. This may create a traffic bottleneck due to the lack of backhaul capacity and unavailability of a prioritization scheme. Thus the quality of voice and other real-time calls degrade significantly. Hence, no prioritization on backhaul link can degrade service. Thus, to guarantee the QoS, there should have SLA and tight inter-operability between operators. A SLA between xDSL, or other cable operator, and the mobile operator can be developed so that both operators are benefitted and the QoS level of femtocell users is ensured. Author in [6] explains that a typical SLA contains the information about, description of the nature of service to be provided, components, the expected performance level of the service, specifically its reliability and responsiveness, the procedure for reporting problems with the service, the time frame for response and problem resolution, the process for monitoring and reporting the service level, the consequences for the service provider not meeting its obligations, and escape clauses and constraints .

Currently almost broadband operators have been deploying network management protocol using the TR-069 specification [7], which is the *Customer Premises Equipment (CPE)* WAN management specification for auto-configuration and remote management. For femtocell deployment, ISP networks should have the capability of prioritizing suitably marked traffics. Therefore, xDSL broadband modems with built-in VoIP ports that automatically prioritize the VoIP traffic above other data traffic for femtocell services. To meet the deployment demands of today, such as IPTV, VoIP, and other services, some operators are considering the TR-098 specification [8] for remote QoS and service differentiation management for future real-time applications. TR-098 provides QoS functionality as well as configuration profiles to ease management and deployment. Currently some ISP operators has been implemented VDSL2 to support IPTV services at home. The TR-098 enables remote configuration of QoS parameters for ISP managed CPEs using an ACS (Auto Configuration Server) (TR-069 server), and is a standard extension to TR-069 protocol. 3GPP uses an inter-domain QoS solution based on SLAs, together with DiffServ marking of User Plane IP packets to exchange QoS information. The *Weighted Fair Queuing (WFQ)*, a flow-based queuing algorithm, can also prioritize low volume femtocell voice traffic. This algorithm schedules low-volume traffic first, while letting high-volume internet and data traffic share the remaining bandwidth.

The BB is an agent responsible for allocating preferred service to users as requested. BB has a policy database to keep the information on who can do what, when and a method of using that database to authenticate requesters. When an allocation is desired for a particular flow, a request is sent to the BB [4]. The requests include the service type, target rate, maximum burst, and the time period when service is required. The BB verifies the unallocated bandwidth whether sufficient to meet the request or not. Under the BB architecture, admission control, resource provisioning, and other policy decisions are performed by a centralized BB in each network domain [5]. BB may handle the xDSL bandwidth for femtocell users efficiently.

## III. Proposed Scheme to Support QoS

The bandwidth satisfaction level of femtocell users' depends on the availability of requested bandwidth. The available bandwidth equal to or greater than requested bandwidth means highest satisfaction level. It means all the requested new calls will be accepted, no dropping or no quality degradation of ongoing calls. But the available bandwidth less than the requested bandwidth means the possibility of some quality degradation or dropping of ongoing calls and blocking of new calls. Hence we define a term "satisfaction level" that depends on the availability of bandwidth. This scheme calculates the amount of reserve bandwidth for femtocell users. The SLA between FAP and xDSL operator will ensure the bandwidth reservation for the femtocell calls. The BB will mange and monitor the distribution of bandwidth among different services.

*A. Bandwidth Reservation Scheme*

Suppose total capacity of xDSL line, required bandwidth by existing internet (non-femtocell users) traffic of house-hold internet applications and bandwidth demand by the femtocell calls are denoted by $C$, $B_I(t)$ and $B_F(t)$ respectively. Then the satisfaction level (*SL*) is calculated as:
Available bandwidth for femtocell calls,

$$B_A(t) = C - B_I(t) \quad (1)$$

if $B_A(t) \geq B_F(t)$, then the satisfaction level,

$$SL(t) = 1 \quad (2)$$

if $B_A(t) < B_F(t)$, then the satisfaction level,

$$SL(t) = \frac{B_A(t)}{B_F(t)} \quad (3)$$

The dynamically reserve bandwidth $B_R(t)$ for the proposed scheme can be calculated as,

$$B_R(t) = \frac{\sum_{n=m}^{m+N} B_F(t - nt_1)}{N} \quad (4)$$

where, $N = \frac{T}{t_1}$ is the total number of observing instants within total observation time period $T$. Within every instants of time interval $t_1$ of total observation time period $T$, the history of requested bandwidth for the femtocell users are considered to calculate the reserve bandwidth at present instant $t$. The value of $n$, $m$, $t_1$ and $T$ can be chosen in the BB by the operators.

if $B_R(t) > B_A(t)$, then the borrowing bandwidth $B_B(t)$ from the existing non-femtocell calls is,

$$B_B(t) = B_R(t) - B_A(t) \quad (5)$$

The SLA between xDSL operator and FAP operator ensure minimum $B_R(t)$ bandwidth for femtocell calls. Hence, available bandwidth for femtocell calls is greater than or equal to $B_R(t)$.

### B. Framework for the Proposed SLA Scheme

Figure 2 shows the framework for our proposed scheme. There is a SLA between the ISP network and femtocell access network to ensure sufficient bandwidth. The BB is the central logical entity that is responsible for the execution of SLA negotiation and resource allocation. The BB knows the current and previous states of bandwidth utilization information. Using this information, BB calculates the future allocation of system bandwidth. The BB monitors the services and their bandwidth allocation, requested bandwidth, type of services, and etc. of the client networks (ISP network and femtocell network). The ISP installs a monitoring scheme to measure the bandwidth allocation among different access networks. The Database stores the information of client networks, SLA policy, and call history. The BB can configure the policy of bandwidth distribution according to the feedback from the monitoring and external input policy. The amount of reserve bandwidth for femtocell users is calculated according to the configuration policy, monitoring result, and collected information from Database. The distribution of bandwidth can be controlled by a rate control device that provides the ability to reserve different bandwidth for different applications. Then the bandwidth is allocated among different services based on SLA.

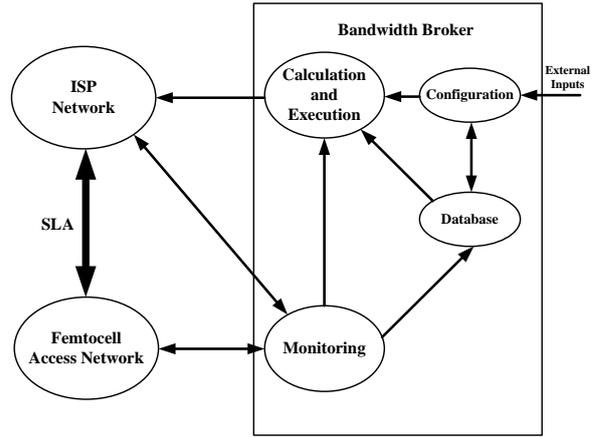

**Figure 2. Framework for the proposed scheme**

## IV. Simulation Result

The performances of the proposed scheme are performed using our simulation results. We performed the satisfaction level of femtocell users that depends on the availability of requested bandwidth. Table 1 shows the basic assumptions and parameters for our simulation. We consider 20 FAPs for our simulation. The parameters are taken by considering WCDMA networks. Then we make an avaerage to get the result for each FAP.

**Table 1. Simulation parameters**

| | |
|---|---|
| xDSL capacity [Mbps] | 6 |
| Non-femtocell traffic arrival | Poisson |
| Femtocell call arrival | Poisson |
| Requested BW by a femtocell voice call [kbps] | 14.4 |
| Requested BW by a femtocell video call [kbps] | 128 |
| Requested BW by a femtocell data call [kbps] | 30 |
| Femtocell voice and video call lifetime [Sec] | 120 |
| Number of femtocell voice call: video call: data call | 5:2:1 |
| Average number of users in the femtocell coverage area | 6 |
| $t_1$ [sec] | 1 |
| T [sec] | 60 |

Figure 3 shows the satisfaction level of femtocell user for traditional scheme (no bandwidth reservation or no prioritization). Different values of average requested bandwidth by nor-femtocell internet traffic (ARBIT) are considered. It shows that, avaerage satisfaction level around 1 can be found when ARBIT = 4.5 Mbps. Even though avaerage required bandwidth for six users is not more than 500 Kbps, satisfaction level is less than 0.9 for ARBIT = 5.5 Mbps. It means that, tarditional scheme can ensure better satisfaction level only for lower internet traffic condition that causes lower bandwidth utilization. Figure 4 shows that the satisfaction level is more than 0.98 for ARBIT = 5.5

Mbps. Even for ARBIT = 6 Mbps, which is equal to the maximum capacity of xDSL, the satisfaction level is more than 0.90. Hence the users can enjoy better quality of services even in heavy internet traffic condition.

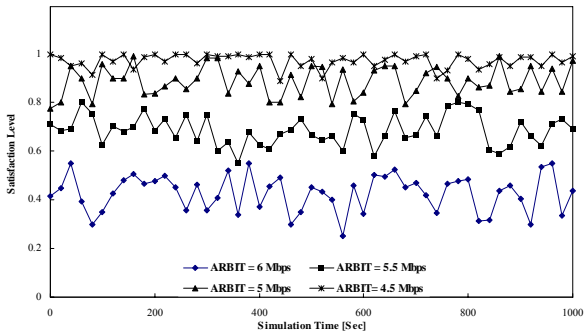

**Figure 3. Femtocell user's satisfaction level in traditional or no prioritization schemes**

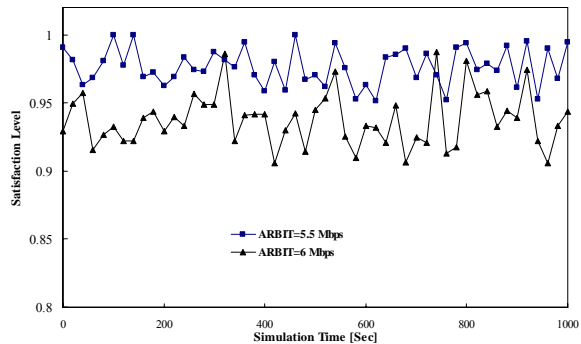

**Figure 4. Femtocell user's satisfaction level with the proposed scheme**

Normally reservation schemes reduce the overall bandwidth utilization of the system. Figure 5 shows that the bandwidth utilization for both the proposed and traditional (no prioritization) schemes are almost same. There is no significant reduction in bandwidth utilization for the proposed bandwidth reservation scheme. But Figure 3 and Figure 4 shows that the proposed scheme increases users' satisfaction level significantly.

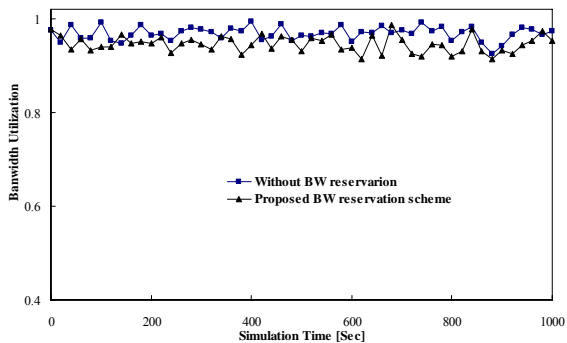

**Figure 5. Comparison between proposed and traditional (no prioritization) schemes in terms of bandwidth utilization for ARBIT=6Mbps**

Figure 6 shows that the bandwidth utilization for the traditional scheme is around 0.75 when ARBIT=4.5 Mbps even though satisfaction level is near to 0.98. But using proposed scheme we can achieve satisfaction level near to 0.98 even ARBIT=5.5 Mbps. Hence the proposed scheme is very much effective for femtocell network deployment.

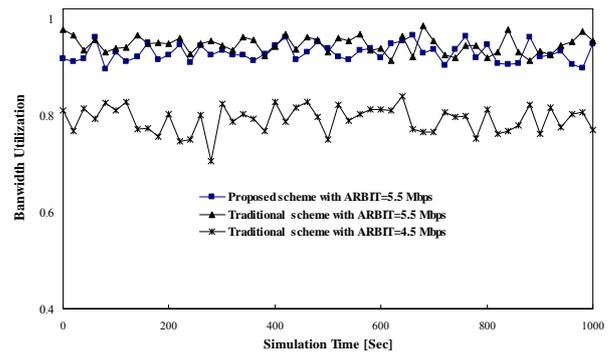

**Figure 6. Comparison between proposed and traditional (no prioritization) schemes in terms of bandwidth utilization to achieve more than 95% satisfaction level**

## V. Conclusion

As femtocell networks mostly handle voice and other real-time services, femtocell backhaul needs to be QoS guaranteed. Femtocells use existing xDSL line for backhauling their traffic. But currently xDSL does not guarantee the QoS for the users. Hence, some new schemes are needed that will prioritize the femtocell calls over other non-femtocell calls. A SLA between xDSL and FAP (mobile operator) operators can ensure the sufficient bandwidth for femtocell users that can potentially resist the degradation of femtocell calls.

From the previous call history, our proposed scheme calculates the approximate bandwidth for upcoming femtocell calls. This amount of bandwidth is reserved for femtocell calls. The simulation results show that our scheme can effectively increase the femtocell users' satisfaction level without significantly decreasing the bandwidth utilization of the overall system. Thus, our approach will be a promising bandwidth allocation technique to ensure the QoS for the real-time services of femtocell networks.